\documentstyle[12pt]{article}
\author{Stefan ~G\"ullenstern \\
Max-Planck-Institut f\"ur Kernphysik,\\
D-69029 Heidelberg, Germany\\
\and Oliver Martin \\
Institut f\"ur Theoretische Physik, Universit\"at Frankfurt\\
D-60054 Frankfurt am Main, Germany
\and Pawe\l\ G\'ornicki \\
Institute of Physics, Polish Academy of Sciences\\
and\\
Center for Theoretical Physics, Polish Academy of Sciences\\
Al. Lotnik\'ow 32/46,
PL-02-668 Warsaw, Poland
\and Lech Mankiewicz \\
Institut f\"ur Theoretische Physik,\\
TU-M\"unchen, Germany
\and Andreas Sch\"afer\\
Institut f\"ur Theoretische Physik, Universit\"at Frankfurt\\
D-60054 Frankfurt am Main, Germany}
\title{{\sc Sphinx v1.1}\\ Monte Carlo Program for Polarized Nucleon-Nucleon
Collisions\\ (UPDATE) }
\begin{document}
\maketitle
\thispagestyle{empty}

\newpage
\noindent
{\LARGE PROGRAM SUMMARY}

\vskip 0.5cm
\begin{trivlist}
\item[\it Title of program:] {\sc Sphinx} -- HEP MONTE CARLO
\item[\it Catalogue number:]
\item[\it Program available from:] Oliver Martin,\\
{\tt http://th.physik.uni-frankfurt.de/$\sim\, $martin/martin/sphinx.html}
\item[\it Computer:] IBM RS6000; PentiumPro based PC, other machines
with {\tt FORTRAN~77} and sufficient capacity should be able to run
this  program
\item[\it Installation:] Institut f\"ur Theoretische Physik,
Goethe-Universit\"at, Frankfurt am Main, Germany
\item[\it Operating system:] tested under {\tt AIX} and {\tt LINUX}
but does not depend
on the particular operating system
\item[\it Programing language used:] {\tt FORTRAN~77}
\item[\it High speed storage required:] 5 Mbytes
\item[\it Number of bits per byte:] 8
\item[\it Number of lines in combined test deck:] 19296
\item[\it CPC subprograms used:] {\sc Jetset~7.3}
\item[\it Keywords:] polarized nucleon-nucleon scattering, high
energy physics, Monte~Carlo simulation
\item[\it Nature of physical problem:]
This program can be used to simulate polarized nucleon - nucleon
collisions at high energies. Spins of colliding particles are
taken into account. The program allows to calculate cross
sections for various processes.
\item[\it Method of solution:]
The existing Monte Carlo program {\sc Pythia~5.6} has been modified to
incorporate spin effects. The program incorporates all features
of {\sc Pythia}.
\item[\it Restrictions on the complexity of the problem:]
Spins of colliding nucleons must be parallel to the collision
axis.
\item[\it Typical running time:] $\approx 0.01$~s on PentiumPro~200
based PC, depends strongly on kinematical cuts
\item[\it References:]
Reader has to refer to {\sc Pythia} manual \cite{pyman} for
additional details.
Further informations can be obtained from:
Oliver Martin, Institut f\"ur Theoretische Physik, Goethe Universit\"at,
Robert-Mayer-Str. 8-10, 60054 Frankfurt am Main, Germany,
e-mail: {\tt martin@th.physik.uni-frankfurt.de} .

\end{trivlist}
\pagebreak

\noindent
{\LARGE LONG WRITE-UP}

\section{Introduction\label{introduction}}
The detailed investigation of hadronic spin effects, most notably in
deep-inelastic lepton-nucleon scattering, has proven to allow for extremly
sensitive tests of QCD. One of the most interesting issues is the possible
contribution of the axial-vector anomaly. This anomaly plays a central role
in modern quantum-field-theory and spin effects might offer the only
possibility to actually detect it. To estimate its contribution one has to
know the distribution function for polarized gluons $\Delta g(x,Q^2)$. It is
argued that the total spin carried by gluons could be large, because
$\alpha_s(Q^2)\Delta g(Q^2)$, with
$\Delta g(Q^2)=\int_0^1{\rm d}x\ \Delta g(x,Q^2)$,
is a renormalisation group invariant, such that if $\Delta g(Q^2)$ is a few
percent at the constituent quark level it could be very sizeable at
large $Q^2$. In deep-inelastic scattering it is impossible to distinguish
between the regular quark contribution and the gluon contribution via the
anomaly. There is actually no conceptually sound way to separate
them. In exclusive or semi-inclusive lepton-hadron scattering
channels it is in principle possible to discern a gluonic and a quark
contribution, but the by far most direct and conceptually clean way are
polarized proton-proton collisions. RHIC \cite{rhic} offers the
option to study
polarized proton collisions up to $\sqrt{s}=300\ {\rm GeV}$ which
would be ideal for this purpose. To analyse this option in detail, to
plan the experiments, and to determine the reachable precission a
detailed Monte Carlo program is needed, which we provide with this
publication. This program can handle, however, only longitudinal
polarization. All polarization effects in proton-proton reactions share
the same problem. As only a small fraction of the partons is polarized
the signal to background ratios are typically of the order percent,
such that a detailed simulation is necessary to extract quantitative results.

Because a state-of-art Monte
Carlo program is usually the result of many years of development
we modified an existing, widly used program
called {\sc Pyhtia} \cite{pyman}. We have been interested
only in those parts of the program that correspond
to nucleon-nucleon collisions. Consequently only these routines were changed.
The resulting polarized version of {\sc Pythia} is named {\sc Sphinx},
an acronym
for {\em {\bf S}imulator of {\bf P}olarised {\bf H}adronic {\bf
IN}teract({\bf X})ions}.

It is interesting to note that {\sc Sphinx} is able to function in unpolarized
mode as well. In that case it is functionally equivalent to {\sc Pythia}.
This means that results will be the same but because some data structures
inside the program have been modified the program requires more
space.

The main problem was to introduce the changes in a way compatible with
the basic structure of {\sc Pythia}, e.\ g.\ the way in which {\sc
Pythia} evaluates the cross sections or simulates the effects of
higher order QCD corrections.
{\sc Pythia} is based on the parton
model. The main assumption is that the nucleon may be represented
by an incoherent mixture of quarks and gluons carrying portions
of nucleon momentum. Their momentum distributions are described by parton
distributions.  In the parton model the nucleon is composed of quarks and
antiquarks of various
flavours and of gluons. In the polarized
case each parton comes in one of two helicities:
righthanded and lefthanded. Therefore the number of parton components
has to be doubled, corresponding to the two degrees of freedom for
longitudinal polarization.
The most natural idea was
to generalize the notion of flavour to account for the polarization.
So instead of d-quark we have lefthanded d-quark and righthanded d-quark
etc. Because in {\sc Pythia} gluons are treated as a special ``flavour''
the scheme
applies to them as well. As far as the general structure of the program
is concerned
the only complication of the polarized code is that we have more
``flavours'' and this is rather easy to implement.

Polarization is followed in {\sc Sphinx} only up to the hard partonic
interaction, i.\ e.\ the hadronic cross section and the higher order
corrections in the initial state shower are evaluated spin{\em dependently}
whereas all final state interactions as fragmentation and
decays are treated spin{\em independently}. No theoretical model for polarized
fragmentation exists. It is,
however, generally accepted that it should not depend on longitudinal
polarization. For transverse polarization the situation would,
however, be different. One knows experimentally that substantial
effects exist, e.\ g.\ a strong coupling between transverse spin and
transverse momentum.

The polarization effects have been implemented in the
following parts of the program:
\begin{enumerate}
\item Parton distributions. At the moment we supply seven sets
of polarized parton distributions: Altarelli-Stirling \cite{altsti}, and
six sets from Gehrmann\&Stirling
\cite{gehrmann}.
First order Altarelli-Parisi \cite{altpar} evolution is taken into
account in all cases.
\item Hard processes. The processes currently implemented in the
polarized mode are summarized in Table \ref{processes}.
Other processes may be used
as well but the results will be averaged over polarizations.
The cross sections for all polarized processes are given in the literature
\cite{crosssec},
and we checked them.
\item Initial state showering. The polarized case was not discussed in
this context
before. We obtained all necessary formulae.
\item Documentation. The event listing has been modified such that it
displays polarization information about the interacting particles up
to the hard partonic interaction.
\end{enumerate}

\begin{table}
\label{processes}
\caption{List of processes implemented in the polarized mode.}
\begin{center}
\begin{tabular}{|r|l|p{8cm}|}
\hline
{\tt ISUB} & Process & Comment \\
\hline\hline
1 & $q_i\bar{q}_j\rightarrow \gamma^* / Z^0 $ & quark-antiquark
annihilation into virtual $ \gamma^* / Z^0$ \\
\hline
2 & $q_i\bar{q}_j\rightarrow W^\pm $ & annihilation into charged
vector boson \\
\hline
11 & $q_iq_j\rightarrow q_iq_j$ &
(anti-)quark -- (anti-)quark scattering; annihilation diagram not
included \\
\hline
12 & $q_i\bar{q}_i\rightarrow q_k\bar{q}_k$ &
annihilation process \\
\hline
13 & $q_i\bar{q}_i\rightarrow g g$ &
annihilation into gluon pair \\
\hline
14 & $q_i\bar{q}_i\rightarrow g \gamma$ &
annihilation into gluon and prompt $\gamma$ \\
\hline
15 & $q_i\bar{q}_i\rightarrow g Z^0$ & annihilation into gluon
and $Z^0$ \\
\hline
16 & $q_i\bar{q}_i\rightarrow g W^\pm$ & annihilation into
gluon and $W^\pm$ \\
\hline
18 & $q_i\bar{q}_i\rightarrow \gamma \gamma$ &
annihilation into $\gamma$ pair \\
\hline
19 & $q_i\bar{q}_i\rightarrow \gamma Z^0$ & annihilation into
$\gamma$ and $Z^0$ \\
\hline
20 & $q_i\bar{q}_i\rightarrow \gamma W^\pm$ & annihilation into
$\gamma$ and $W^\pm$ \\
\hline
28 & $q_ig\rightarrow q_ig$ &
(anti-)quark -- gluon scattering \\
\hline
29 & $q_ig\rightarrow q_i\gamma$ &
prompt $\gamma$ production in (anti-)quark -- gluon scattering \\
\hline
30 & $ q_ig \rightarrow q_iZ^0$ & $Z^0$ production in
(anti-)quark -- gluon scattering \\
\hline
31 & $ q_ig \rightarrow q_jW^\pm$ & $W^\pm$ production in
(anti-)quark -- gluon scattering \\
\hline
53 & $g g\rightarrow q_k\bar{q}_k$ &
gluon fusion \\
\hline
68 & $g g\rightarrow g g$ &
gluon -- gluon scattering \\
\hline \hline
\end{tabular}
\end{center}
\end{table}

Hereafter we do not try to explain the structure of {\sc Pythia~5.6}
and we indicate
only the most significant modifications which lead to {\sc Sphinx}.
Readers unfamiliar with {\sc Pythia}
should first consult the appropriate manual \cite{pyman}.
\section{From {\sc Pythia} to {\sc Sphinx}: modifications\label{main}}
Our program is an extension of {\sc Pythia} version 5.6.
The {\sc Pythia~5.6} program heavily relies on {\sc Jetset~7.3}
subroutines \cite{jetset}.
These subroutines are mainly used to simulate processes that take
place after the hard partonic interaction, i.e. fragmentation and
decays, or provide spinindependent manipulations as for example
Lorentz-transformations.
Because polarization
effects in the final state are neglected in {\sc Sphinx},
the {\sc Jetset} program could be left untouched. Only the set-up of an event
listing which is also a task of {\sc Jetset} in a {\sc Pythia} run has
been taken over from {\sc Sphinx} in the polarized mode. The
event listing in {\sc Sphinx} provides then in addition information
about the polarization flow.
For
that purpose two new subroutines have been added which are
modifications of the corresponding {\sc Jetset} routines (see below).
By this means {\sc Sphinx} provides a suitable interface to {\sc
Jetset} such that the programs
may be linked without any modifications in the latter.
There is only one exception: the {\sc Jetset} subroutine {\tt LUGIVE},
which returns the values of all {\sc Jetset} {\em and} {\sc Pythia}
common block variables cannot be longer used, because in {\sc Sphinx}
the dimensions of some {\sc Pythia} arrays have been enlarged so that
the formats no longer match.

In many places inside {\sc Pythia} partonic data are stored in arrays indexed
by flavour. Inclusion of helicities means doubling the number of ``flavours''
as described in Section \ref{introduction}.
For that reason we had to enlarge the arrays.
The unpolarized mode has been preserved throughout the program
and works exactly as the original
code.
All modifications are implemented in a treelike structure with several
branching points where the user may choose between the
polarized and unpolarized modes. The default is always unpolarized.
An interesting point is that one can combine polarized and unpolarized
treatments for different processes and for various stages  of the same
process. This has to be done with care and one has to be aware of this
not always being justified from the
physical point of view.
The branching points were
implemented by means of the {\tt IF-ELSE} structure
which directs the program flow according to a specific switch.
Each functional part (usually a subroutine) has its own {\em local}
variable
called {\tt IPOL}. This variable controls the mode and its value depends
on some
parameters\footnote{See Table \ref{parameters}} {\tt MSTP(x)} and {\tt
NSUB(x)}
to be defined by the user in the main program.
For example, it is
possible to run the code with polarized partonic cross
section and polarized parton distributions but unpolarized initial state
showering etc. This solution provides flexibility but has to be
used carefully.
The major disadvantage of {\tt IF-ELSE} constructions is that parts of
the code are multiplied. This makes servicing
of the code more cumbersome because one has to introduce the same
change in different places at once. In addition the resulting code
became quite lengthy.

{\sc Sphinx} is as {\sc Pythia} a ``slave-system'', i.\ e.\ it
consists only of callable subroutines where two of them ({\tt
PYINIT} and {\tt PYEVNT}) have to be called by the user to
perform the event generation and a few other, e.\ g.\ {\tt DPLIST},
{\tt PYSTAT}, etc., could be called to obtain further event information.
The rest of the subroutines is of internal use. Therefore the user has
to supply
a main program where all relevant parameters and switches have to be
specified and these subroutines have to be called. Because this
structure is the same as in {\sc Pythia} the reader is referred to
\cite{pyman} again for details about the general {\sc Pythia}
parameters. In this article only the new parameters in {\sc Sphinx}
are discussed.
The modifications made in the individual subroutines are described in Section
\ref{subroutines}.
In Table \ref{parameters} we list all new parameters introduced
to control the polarized mode.
Finally in Section \ref{example} examples for a main program
and the result of the corresponding test runs are presented.
\begin{table}
\label{parameters}
\caption{Parameters controlling the polarized mode}
\begin{center}
\begin{tabular}{|l|rp{8cm}|c|c}
\hline
Parameter &&Description & Default \\
\hline\hline
{\tt MSTP(171)} && beam polarization & 0\\
&=0:& unpolarized  & \\
&=1:& polarization in $+z$ direction &\\
&=2:& polarization in $-z$ direction &\\
\hline
{\tt MSTP(172)} && target polarization & 0\\
&=0:& unpolarized  & \\
&=1:& polarization in $+z$ direction &\\
&=2:& polarization in $-z$ direction &\\
\hline
{\tt MSTP(175)} && use of polarized parton distribution in polarized
initial state shower ({\tt MSTP(176)=1})& 1\\
&=0:& unpolarized distribution; for testing only, {\em do not use!}&\\
&=1:& polarized distribution&\\
\hline
{\tt MSTP(176)} && initial state showering mode & 0\\
&=0:& unpolarized & \\
&=1:& polarized & \\
\hline
\end{tabular}
\end{center}
\end{table}

\setcounter{table}{1}
\begin{table}
\caption{Parameters controlling the polarized mode, continued}
\begin{center}
\begin{tabular}{|l|rp{8cm}|c|c}
\hline
Parameter &&Description & Default \\
\hline\hline
{\tt MSTP(177)} && set of polarized parton distributions
$\Delta q$ and $\Delta g$ used; in addition one has
to specify an unpolarized set as in standard {\sc Pythia} & 0\\
&=0:& $\Delta q =0$ and $\Delta g=0$ (no polarization) & \\
&=1:& fake polarization, built up from unpolarized distribution
according to $\Delta q = \frac{\tt MSTP(178)}{100}q$ & \\
&=2:& Altarelli-Stirling parametrization \cite{altsti}; data
file {\tt altsti.dat} required & \\
&=3:& Gehrmann-Stirling parametrization \cite{gehrmann} lo~set~a; data
file {\tt gestloa.dat} required & \\
&=4:& Gehrmann-Stirling parametrization \cite{gehrmann} lo~set~b; data
file {\tt gestlob.dat} required & \\
&=5:& Gehrmann-Stirling parametrization \cite{gehrmann} lo~set~c; data
file {\tt gestloc.dat} required & \\
&=6:& Gehrmann-Stirling parametrization \cite{gehrmann} nlo set~a; data
file {\tt gestnloa.dat} required & \\
&=7:& Gehrmann-Stirling parametrization \cite{gehrmann} nlo set~b; data
file {\tt gestnlob.dat} required & \\
&=8:& Gehrmann-Stirling parametrization \cite{gehrmann} nlo set~c; data
file {\tt gestnloc.dat} required & \\
&=9:& fake polarization, for testing only, {\em do not use!}& \\
\hline
\end{tabular}
\end{center}
\end{table}

\setcounter{table}{1}
\begin{table}
\caption{Parameters controlling the polarized mode, continued}
\begin{center}
\begin{tabular}{|l|rp{8cm}|c|c}
\hline
Parameter &&Description & Default \\
\hline\hline
{\tt MSTP(178)}&& percentage of fake polarization for {\tt MSTP(177)=1} & 0\\
\hline
{\tt MSTP(180)} && mode selection (master switch) & 0\\
&=0:& unpolarized mode; this value overrides all other
polarization switches & \\
&=1:& polarized mode  & \\
\hline
{\tt NSUB(ISUB)} && mode for subprocess ISUB & 0 \\
&=0:& unpolarized treatment &\\
&=1:& polarized treatment &\\
\hline
{\tt NSEL} && menue of polarized processes & 0 \\
&=1:&  {\tt ISUB = 11,12,13,28,53,68} switched on&\\
&=10:& {\tt ISUB = 14,18,29} switched on&\\
&=11:& {\tt ISUB = 1} switched on&\\
&=12:& {\tt ISUB = 2} switched on &\\
&=13:& {\tt ISUB = 15,30} switched on &\\
&=14:& {\tt ISUB = 16,31} switched on&\\
\hline
\end{tabular}
\end{center}
\end{table}
\begin{table}
\label{internal}
\caption{Internal variables storing polarization information}
\begin{center}
\begin{tabular}{|l|rp{8cm}|c|c}
\hline
Variable &&Description & Com.\ Block \\
\hline\hline
{\tt MINT(311)} && beam helicity & {\tt PYINT1}\\
&=0:& unpolarized  & \\
&=1:& positive helicity &\\
&=2:& negative helicity&\\
\hline
{\tt MINT(312)} && target helicity & {\tt PYINT1}\\
&=0:& unpolarized  & \\
&=1:& positive helicity &\\
&=2:& negative helicity&\\
\hline
{\tt MINT(313)} && helicity of shower initiator on beam side& {\tt PYINT1}\\
&=0:& unpolarized  & \\
&=1:& positive helicity &\\
&=2:& negative helicity&\\
\hline
{\tt MINT(314)} && helicity of shower initiator on target side& {\tt PYINT1}\\
&=0:& unpolarized  & \\
&=1:& positive helicity &\\
&=2:& negative helicity&\\
\hline
{\tt MINT(315)} && helicity of hard interacting parton on beam side&
{\tt PYINT1}\\
&=0:& unpolarized  & \\
&=1:& positive helicity &\\
&=2:& negative helicity&\\
\hline
{\tt MINT(316)} && helicity of hard interacting parton on target side&
{\tt PYINT1}\\
&=0:& unpolarized  & \\
&=1:& positive helicity &\\
&=2:& negative helicity&\\
\hline
{\tt MSTP(179)}&& switch off polarization temporarely in {\tt PYSIGH}
and {\tt PYSTFU} resp. & {\tt PYPARS}\\
&=0:& no action   & \\
&=1:& switch off polarization &\\
\hline
\end{tabular}
\end{center}
\end{table}
\setcounter{table}{2}
\begin{table}
\caption{Internal variables storing polarization information, continued}
\begin{center}
\begin{tabular}{|l|rp{7cm}|c|c}
\hline
Variable &&Description & Com.\ Block \\
\hline\hline
{\tt ISIGH(1000,6)} && hard scattering information of {\tt I}th line & {\tt
PYINT3}\\
{\tt ISIGH(I,1)}&& particle code of {\tt I}th line on beam side  & \\
{\tt ISIGH(I,2)}&& particle code of {\tt I}th line on target side  & \\
{\tt ISIGH(I,3)}&& colour flow  & \\
{\tt ISIGH(I,4)}&& helicity of {\tt I}th line on beam side  & \\
{\tt ISIGH(I,5)}&& helicity of {\tt I}th line on target side  & \\
{\tt ISIGH(I,6)}&& not used  & \\
\hline
{\tt KD(I)} && polarization/helicity of {\tt I}th line & {\tt DPYPOL}\\
&=0:& no polarization/helicity   & \\
&=1:& positive polarization/helicity   & \\
&=2:& negative polarization/helicity   & \\
\hline
{\tt XSFX(2,-40:40,0:2)} && $x$ times parton distribution for given $x$
and $Q^2$ of flavour {\tt KFL = -40:40} and helicity {\tt KFLD = 0:2} on beam
side ({\tt JT=1}) and target side
({\tt JT=2}) resp.& {\tt PYINT3}\\
{\tt XSFX(JT,KFL,0)}&& unpolarized   & \\
{\tt XSFX(JT,KFL,1)}&& positive helicity   & \\
{\tt XSFX(JT,KFL,2)}&& negative helicity   & \\
\hline
\end{tabular}
\end{center}
\end{table}

\begin{table}
\caption{\label{tabup}
Parameters controlling the choice of the unpolarized parton
distribution functions}
\begin{center}
\begin{tabular}{|l|rp{8cm}|c|c}
\hline
Parameter &&Description & Default \\
\hline\hline
{\tt MSTP(51)} && set of unpolarized parton distributions;
see also {\tt MSTP(52)} & 12\\
&=1:& EHLQ1 set (1986 updated version, LO)& \\
&=2:& EHLQ2 set (1986 updated version, LO)& \\
&=3:& Duke-Owens set 1 (LO)&\\
&=4:& Duke-Owens set 2 (LO)&\\
&=5:& CTEQ2M (NLO)& \\
&=6:& CTEQ2MS (NLO)& \\
&=7:& CTEQ2MF (NLO)& \\
&=8:& CTEQ2ML (NLO)& \\
&=9:& CTEQ2L (LO)& \\
&=10:& CTEQ2D (NLO)& \\
&=11:& Gl\"uck, Reya, Vogt (LO) & \\
&=12:& Gl\"uck, Reya, Vogt 94 (LO) & \\
&=13:& Martin, Roberts, Stirling MRS(AP) (NLO) & \\
&=14:& Martin, Roberts, Stirling MRS(R1) (NLO),
{\em has not been tested yet!} & \\
&=15:& Martin, Roberts, Stirling MRS(R2) (NLO),
{\em has not been tested yet!} & \\
\hline
{\tt MSTP(52)} && choice of proton parton-distribution-library & 1\\
&=1:& the internal {\sc Sphinx} one, with parton distributions according
to {\tt MSTP(51)}above & \\
&=2:& the {\sc Pdflib} one, with the {\sc Pdflib} (version~4 and higher)
{\tt NGROUP} and {\tt NSET} numbers to be given as
{\tt MSTP(51)=1000*NGROUP+NSET}. In order to make use of this option
{\sc Pdflib} must be linked and the subroutines {\tt STRUCTM} and
{\tt PDFSET} at the end of the program should be commented out. & \\
\hline
\end{tabular}
\end{center}
\end{table}

\section{Common Blocks and Subroutines\label{subroutines}}
In the following the modified and the new
subroutines that have been specifically created for {\sc Sphinx} are
described in more detail. The general tasks of the subroutines
themselves as well as the unchanged parameters and variables are not
explained, because they are the same as in {\sc Pythia}. The reader is
asked again if occasion arises to consult \cite{pyman} to obtain the
needed information. We restrict our explanation to the new aspects in
{\sc Sphinx}.
The purpose of the modifications is
indicated and the new parameters, switches, and internal variables are
listed. Meaning and possible values of the new parameters are
given in Table \ref{parameters}.
To incorporate polarization the following common blocks have been
enlarged and replace the
corresponding {\sc Pythia} common blocks or are added:
\begin{itemize}
\item {\tt COMMON/PYINT3/XSFX(2,-40:40,0:2),ISIG(1000,6),SIGH(1000)}
\item {\tt
COMMON/PYSUBS/MSEL,NSEL,MSUB(200),NSUB(200),KFIN(2,-40:40),
CKIN(200)}
\item {\tt COMMON/DPYPOL/KD(4000)}
\end{itemize}
Information about the new
internal variables and enlarged arrays can be found in Table \ref{internal}.
In addition
it is shown how the local polarization switch {\tt IPOL} is built up
in the different subroutines. Only the
polarized case ({\tt IPOL=1}) will be discussed, because in the
unpolarized case ({\tt IPOL=0}) each subroutine works exactly as the
corresponding {\sc Pythia} subroutine.
The not mentioned subroutines of {\sc Sphinx} are the same as in {\sc Pythia}.

\paragraph{\tt MAIN PROGRAM}\hfill\break
\noindent{\bf Purpose:} to set up the polarized event generation. The
variables
which have to been set are listed in Table \ref{parameters}.\hfill\break
{\bf Remarks:} Examples of a main program are given in Section
\ref{example}.

\paragraph{\tt SUBROUTINE PYINIT(FRAME,BEAM,TARGET,WIN)}\hfill\break
\noindent{\bf Purpose:} to display {\sc Sphinx} header; to check
partially the availability of the
desired polarization
scenario, i.\ e.\ to control that the master switch for polarization
{\tt MSTP(180)} is set properly, the selected partonic
subprocesses can be treated polarized and to control and compose the
polarization
menue via {\tt NSEL}; to call {\tt DPLIST} instead of {\tt
LULIST} (see below).\hfill\break
{\bf New parameters:} {\tt MSTP(180)}, {\tt NSEL}, {\tt
NSUB(ISUB)}\hfill\break
{\bf Internal Polarization switch:} {\tt IPOL=MSTP(180)}\hfill\break
{\bf Remarks:} The interface to the CERN parton distribution library
was updated and works now with {\sc Pdflib~4} and higher versions.
Furthermore, if a not allowed scenario has been chosen, the
programs stops with an appropriate error message.
The {\sc Pakpdf} parton distribution library is not supported anymore.

\paragraph{\tt SUBROUTINE PYEVNT}\hfill\break
\noindent{\bf Purpose:} to start polarized event generation;
to call {\tt DPEDIT} instead of {\tt LUEDIT} (see below).\hfill\break
{\bf New parameters:} {\tt MSTP(180)}\hfill\break
{\bf Internal polarization switch:} {\tt IPOL=MSTP(180)}

\paragraph{\tt SUBROUTINE PYINKI(CHFRAM,CHBEAM,CHTARG,WIN)}\hfill\break
\noindent{\bf Purpose:} to check availability of the desired hadronic
polarization
scenario, i.\ e.\ to control that the selected hadron
can be treated polarized and to verify that the polarization
is longitudinal; to store the polarization of beam and target for the
event listing in {\tt KD(1)} and {\tt KD(2)} and for internal use in
{\tt MINT(311)} and {\tt MINT(312)}.\hfill\break
{\bf New parameters:} {\tt KD(I)}, {\tt MSTP(171)}, {\tt MSTP(172)},
{\tt MSTP(180)} \hfill\break
{\bf New internal variables:} {\tt MINT(311)}, {\tt MINT(312)}\hfill\break
{\bf Internal polarization switch:} {\tt IPOL=MSTP(180)}\hfill\break
{\bf Remarks:} At present only nucleons and hyperons and their
antiparticles can be treated polarized.
If a not allowed scenario has been chosen, the
programs stops with an appropriate error message.

\paragraph{\tt SUBROUTINE PYRAND}\hfill\break
\noindent{\bf Purpose:} to adapt {\tt PYRAND} to the new
environment -- all relevant arrays which have been enlarged or added to the
common blocks in other subroutines are modified here as well; to
extend event shape selection to incorporate helicities;
to store
helicities of the partons entering the hard interaction according to
\begin{itemize}
\item {\tt MINT(313)}: helicity of the beam parton for use in the initial
state showering subroutine;
\item {\tt MINT(314)}: helicity of the target parton for use in the initial
state showering subroutine;
\item {\tt MINT(315)}: helicity of the beam parton;
\item {\tt MINT(316)}: helicity of the target parton.
\end{itemize}
\hfill\break
{\bf New parameters:} {\tt MSTP(180)}, {\tt NSUB(ISUB)} \hfill\break
{\bf New internal variables:} {\tt MINT(313)}, {\tt MINT(314)}, {\tt
MINT(315)}, {\tt MINT(316)}\hfill\break
{\bf Internal polarization switch:} {\tt IPOL=MSTP(180)$\times$NSUB(ISUB)}
\hfill\break
{\bf Remarks:}
Note that {\tt MINT(313)=MINT(315)} and {\tt MINT(314)=MINT(316)} but the
values of {\tt MINT(313)} and {\tt MINT(314)} are changed later by
the initial state shower in {\tt PYSSPA}.

\paragraph{\tt SUBROUTINE PYSCAT}\hfill\break
\noindent{\bf Purpose:} to adopt {\tt PYSCAT} to the new environment
(see {\tt PYRAND}); to store helicities of the partons entering the
hard interaction; to fill lines 1, 2 and 5, 6 in the event listing
with polarization information (see below).
\hfill\break
{\bf New parameters:} {\tt KD(I)}, {\tt MSTP(180)}, {\tt NSUB(ISUB)}
\hfill\break
{\bf New internal variables:} {\tt MINT(315)}, {\tt MINT(316)}\hfill\break
{\bf Internal polarization switch:} {\tt
IPOL=MSTP(180)$\times$NSUB(ISUB)}

\paragraph{\tt SUBROUTINE PYSSPA(IPU1,IPU2)}\hfill\break
\noindent{\bf Purpose:} to perform polarized initial state
showering, helicity dependent GLAP evolution equations are used in the
backward evolution algorithmus; to enlarge all relevant array in an
appropriate manner to incorporate polarization;
to check proper selection of the
polarized initial state shower
scenario, i.\ e.\ to control that the {\tt MSTP(175)} and {\tt
MSTP(176)} are set correctly; to store the helicities of initial state
shower initiators ({\tt MINT(313)}, {\tt MINT(314)}).\hfill\break
{\bf New parameters:} {\tt KD(I)}, {\tt MSTP(171)}, {\tt MSTP(172)},
{\tt MSTP(175}, {\tt MSTP(176)}, {\tt MSTP(180)}, {\tt NSUB(ISUB)}
\hfill\break
{\bf New internal variables:} {\tt MINT(313)}, {\tt MINT(314)}, {\tt
MINT(315)}, {\tt MINT(316)}\hfill\break
{\bf Internal polarization switch:}
{\tt IPOL=MSTP(180)$\times$NSUB(ISUB)$\times$MSTP(176)}
{\bf Remarks:} At the present stage only QCD shower can be treated
polarized, QED showering has to be done in the unpolarized manner.
The combination {\tt MSTP(175)=0} and {\tt MSTP(176)=1} allows to
simulate {\em polarized} showering with the use of {\em unpolarized}
parton distributions. This option is just for testing and should not
be selected by the user!
If {\tt MSTP(175)} or {\tt MSTP(176)} are set improperly, the
programs stops with an appropriate error message. The internal
variables
{\tt MINT(313)} and {\tt MINT(314)} are changed to their final values
in this subroutine.

\paragraph{\tt SUBROUTINE PYMULT(MMUL)}\hfill\break
\noindent{\bf Purpose:} to switch off polarization in {\tt PYSIGH}
(set {\tt MSTP(179)=1} temporally)
when called from {\tt PYMULT} even in a polarized run, because
multiple interaction cannot be treated polarized at the moment.
\hfill\break
{\bf New parameters:} {\tt MSTP(179)}

\paragraph{\tt SUBROUTINE PYREMN(IPU1,IPU2)}\hfill\break
\noindent{\bf Purpose:} to adopt {\tt PYREMN} to the new environment
(see {\tt PYRAND}); to fill lines 3, 4 in the event listing
with polarization information (see below).
\hfill\break
{\bf New parameters:} {\tt KD(I)}

\paragraph{\tt SUBROUTINE PYSIGH}\hfill\break
\noindent{\bf Purpose:} to evaluate the helicity dependent hadronic
cross sections by convolution of the helicity dependent parton
distributions with the helicity dependent partonic cross sections; to
supply the subroutine with the helicity dependent partonic cross
sections.\hfill\break
{\bf New parameters:} {\tt MSTP(171)}, {\tt MSTP(172)}, {\tt
MSTP(179)}, {\tt MSTP(180)}, \hfill\break{\tt
NSUB(ISUB)}\hfill\break
{\bf Internal Polarization switch:} \hfill\break{\tt
IPOL=MSTP(180)$\times$NSUB(ISUB)$\times$(1-MSTP(179))}\hfill\break
{\bf Remarks:} {\tt PYSIGH} will always run in the unpolarized mode
when it is called by {\tt PYMULT} which sets {\tt MSTP(179)=1} in {\tt
PYSIGH} temporally. Evaluating the spindependent hadronic cross
sections one has to notice that the hadrons are specified according to
the spin, whereas the partons are labelled by their helicities.
The spin is defined relative to the
collision axis and the beam is assumed to move in the positive
direction. For that reason the helicities at the target side are
opposite to the polarizations. Hence the target labels are reversed in
the convolution in comparison to the beam labels.
The parton distributions are passed from {\tt PYSTFU} to {\tt PYSIGH}
through the array
{\tt XPQ(KFL,KFLD)} (see below) and stored in the array
{\tt XSFX(N,KFL,KFLD)}, where {\tt KFLD} denotes the helicity.
{\tt ISIG(N,I)} contains the information about the Nth-line in the
event listing. The new entries {\tt ISIG(N,4)} and {\tt ISIG(N,5)}
store the helicities of the partons at the beam and target side
respectively. {\tt ISIG(N,6)} is reserved but not used at the moment.

\paragraph{\tt SUBROUTINE PYSTFU(KF,X,Q2,XPQ)}\hfill\break
\noindent{\bf Purpose:} to evaluate the helicity dependent parton
distributions for given flavour ({\tt KF}), x ({\tt X}), and $Q^2$
({\tt Q2}) according to the selected parametrizations.\hfill\break
{\bf New parameters:} {\tt MSTP(177)}, {\tt MSTP(178)}, {\tt
MSTP(179)}, {\tt MSTP(180)}, \hfill\break{\tt
NSUB(ISUB)}\hfill\break
{\bf Internal Polarization switch:} {\tt
IPOL=MSTP(180)$\times$(1-MSTP(179))}\hfill\break
{\bf Remarks:}
Call of {\tt PYSTFU} returns x times the parton distribution functions for
given flavour, $x$,  and $Q^2$ for both helicities and an averaged
(unpolarized) value.
The values are stored in the
array {\tt XPQ(KFL,KFLD)} which has been enlarged from {\tt XPQ(-25:25)} to
{\tt XPQ(-25:25,0:2)}. Row {\tt XPQ(KFL,0)}
contains the unpolarized distributions, {\tt XPQ(KFL,1)} distributions
corresponding to the positive
helicity (relative to the hadron) and {\tt XPQ(KFL,2)} distributions for
the negative helicity.
The parton distributions are selected by switches described earlier (see Table
\ref{parameters}). The polarized distributions $q_\pm$, $g_\pm$ are
constructed from the
unpolarized ones $q$, $g$ (selected by old {\sc Pythia} switches) and
polarized parts
$\Delta q$ and $\Delta g$ selected by {\tt MSTP(177)} (see Table
\ref{parameters}). Four subroutines have been added to calculate
the polarized distributions. These are
{\tt ALTSTI}\footnote{This subroutine has been written by G.\
Altarelli and J.\ Stirling.
Used with permision from the authors.},
{\tt POLLO}, and {\tt POLNLO}\footnote{These subroutines have
 been written by T.\ Gehrmann and W.J.\ Stirling.
Used with permision from the authors.}
The subroutines require data
files {\tt altsti.dat}, {\tt gestloa.dat}, {\tt gestlob.dat},
{\tt gestloc.dat}, {\tt gestnloa.dat}, {\tt gestnlob.dat}, and
{\tt gestnloc.dat}. These
files must
be visible to {\tt FORTRAN} {\tt open} statement and therefore
they have to be placed in an appropriate directory.
The files are supplied with the program.
The parton distribution for fixed helicity are reconstructed from the
polarized and unpolarized distributions according to
$q_\pm=\frac{1}{2}\left(q\pm\Delta q\right)$.
When the polarized and unpolarized parts are combined together
the program performs the unitarity check -- if the resulting total
distribution
becomes negative for one helicity it is put to zero and the
corresponding result for the other helicity is set to the value of the
unpolarized part.
Only polarized parametrizations for protons are implemented. Neutron
parametrizations are obtained from them by isospin symmetry, the
parametrizations for hyperons are constructed by naive SU(3). Charge
conjugations is used to describe the corresponding antiparticles.

The interface to the CERN parton distribution library was updated and
works now with {\sc Pdflib~4} and higher versions.
The {\sc Pakpdf} parton distribution library is not supported anymore.
Furhermore, several new unpolarized parton distribution sets have been
included in {\sc Sphinx} (see Table~\ref{tabup}).

{\em
When plotting ratios of polarized and unpolarized parton distribution
functions
one should have in mind that the distributions are often
calculated from a two-dimensional grid by means of linear interpolation.
These grids are usually logarithmic in $x$ and $Q^2$. Furthermore, the
subroutines which give the polarized and unpolarized distributions
may use different $(x,Q^2)$ points and different interpolation routines.
As a result, the asymmetry may not be smooth
in the large-$x$ region. Usually, this problem
can be circumvented by using a parametrization of unpolarized
parton distributions which is rather given by a fit with polynomials
than by interpolation between data points. This method is usually slower
but provides much smoother results. In general it is recommended to use the
set of unpolarized parton distributions which corresponds to the polarized
set.
For the Gehrmann/Stirling parametrizations
 these are the GRV94(LO) and MRS(AP)(NLO) sets,
respectively. These sets are available in {\sc Sphinx}
as fits with polynomials by setting
{\tt MSTP(51)=12} and {\tt MSTP(51)=13}, respectively.
}

\paragraph{\tt SUBROUTINE PYSTPR(X,Q2,XPPR)}
\hfill\break
\noindent{\bf Purpose:}
to calculate the unpolarized parton distributions of the proton;
\hfill\break
\noindent{\bf Remark:}
the implementation of the {\sc Cteq2} type unpolarized parton distributions
was updated.
The Interface to {\tt PYCTQ2} was taken from {\sc Pythia~5.7}
Several new unpolarized sets were included (see Table~\ref{tabup}).

\paragraph{\tt FUNCTION PYCTQ2 (Iset,Iprt,X,Q)}
\hfill\break
\noindent{\bf Purpose:}
to give the revised {\sc Cteq2} parton distribution sets
with extended range in
parametrized form.
\hfill\break
\noindent{\bf Remark:} This function was taken from {\sc Pythia~5.7}.

\paragraph{\tt  SUBROUTINE MRSAP(X,SCALE,UPV,DNV,USEA,DSEA,STR,CHM,BOT,GLU)}
\hfill\break
\noindent{\bf Purpose:}
to calculate the unpolarized parton distribution functions set~A'
from Martin, Roberts and Stirling. $x$ and $Q$ are passed to the subroutine
via the parameters {\tt X} and {\tt SCALE}.
{\tt UPV}, {\tt DNV}, {\tt USEA}, {\tt DSEA},
{\tt STR}, {\tt CHM}, {\tt BOT}, and {\tt GLU} contain
$u_{\rm val}$, $d_{\rm val}$, $\bar u$, $\bar d$,
$\bar s$, $\bar c$, $\bar b$, and $g$ repectively.
\hfill\break
\noindent{\bf Remarks:} This subroutine was written by A.D.~Martin,
R.G.~Roberts, and W.J.~Stirling. For further information see~\cite{mrsap}.

\paragraph{\tt  SUBROUTINE MRSR1(X,Q2,UPV,DNV,USEA,DSEA,STR,CHM,BOT,GLU)}
\hfill\break
\noindent{\bf Purpose:}
to return the unpolarized parton distribution functions set~R1
from Martin, Roberts and Stirling. $x$ and $Q^2$ are passed to the subroutine
via the parameters {\tt X} and {\tt Q2}.
{\tt UPV}, {\tt DNV}, {\tt USEA}, {\tt DSEA},
{\tt STR}, {\tt CHM}, {\tt BOT}, and {\tt GLU} contain
$u_{\rm val}$, $d_{\rm val}$, $\bar u$, $\bar d$,
$\bar s$, $\bar c$, $\bar b$, and $g$ repectively.
\hfill\break
\noindent{\bf Remarks:} This subroutine was written by A.D.~Martin,
R.G.~Roberts, and W.J.~Stirling. For further information see~\cite{mrsr}.

\paragraph{\tt  SUBROUTINE MRSR2(X,Q2,UPV,DNV,USEA,DSEA,STR,CHM,BOT,GLU)}
\hfill\break
\noindent{\bf Purpose:}
to calculate the unpolarized parton distribution functions set~R2
from Martin, Roberts and Stirling. $x$ and $Q^2$ are passed to the subroutine
via the parameters {\tt X} and {\tt Q2}.
{\tt UPV}, {\tt DNV}, {\tt USEA}, {\tt DSEA},
{\tt STR}, {\tt CHM}, {\tt BOT}, and {\tt GLU} contain
$u_{\rm val}$, $d_{\rm val}$, $\bar u$, $\bar d$,
$\bar s$, $\bar c$, $\bar b$ and $g$ repectively.
\hfill\break
\noindent{\bf Remarks:} This subroutine was written by A.D.~Martin,
R.G.~Roberts, and W.J.~Stirling. For further information see~\cite{mrsr}.

\paragraph{\tt SUBROUTINE GRV94(X,Q2,UV,DV,DEL,UDB,SB,GL)}
\hfill\break
\noindent {\bf Purpose:}
to return the leading order (1994) unpolarized parton distribution functions
from Gl\"uck, Reya, Vogt. $x$ and $Q^2$ are passed to the subroutine
via the parameters {\tt X} and {\tt Q2}.
{\tt UV}, {\tt DV}, {\tt DEL}, {\tt UDB},
{\tt SB}, and {\tt GL} contain
$u_{\rm val}$, $d_{\rm val}$, $\bar d - \bar u$, $\bar d + \bar u$,
$\bar s$, and $g$ respectively. The heavy quark distributions
$c$, $b$, and $t$ are zero in the whole $Q^2$-range.
\hfill\break
\noindent{\bf Remarks:} This subroutine was written by M.~Gl\"uck,
E.~Reya, and A.~Vogt. For further information see \cite{grv94}.


\paragraph{\tt SUBROUTINE DPLIST(MLIST)}\hfill\break
\noindent{\bf Purpose:} to display the polarizations of the particle
in the event listing.\hfill\break
{\bf New parameters:} {\tt KD(I)}
\hfill\break
{\bf Remarks:} {\tt DPLIST} is a modification of the {\sc Jetset}
subroutine {\tt LULIST}.
It is changed to display the
polarization in the final listing. The sign displayed just
behind the particle code denotes polarization with respect to
the z-axis. When the sign is
missing the particle has been treated as unpolarized.
The information is taken from the vector {\tt KD(I)} and transformed
accordingly to (`$0$',`$1$',`$2$')$\rightarrow$(`$\ $',`$+$',`$-$').
The following format is chosen (polarization for
the colliding hadrons, helicity for the partons resp.):
\begin{itemize}
\item = `  ': no polarization/helicity\\
\item = `$+$': positive polarization/helicity\\
\item = `$-$': negative polarization/helicity
\end{itemize}

\paragraph{\tt SUBROUTINE DPEDIT(MEDIT)}\hfill\break
\noindent{\bf Purpose:} to compress the vector {\tt KD(I)}, containing
the polarization information,
properly.\hfill\break
{\bf New parameters:} {\tt KD(I)}
\hfill\break
{\bf Remarks:} {\tt DPEDIT} is a modification of the {\sc Jetset}
subroutine {\tt LUEDIT}.

\paragraph{\tt SUBROUTINE
ALTSTI(X,Q2,UPV,DNV,SEA,STR,CHM,BOT,TOP,GLU)}\hfill\break
\noindent{\bf Purpose:} to return $x$ times the polarized parton
distributions evaluated at given $x$ ({\tt X}) and $Q^2$ ({\tt Q2})
according to the parametrization of Altarelli\&Stir\-ling \cite{altsti}.
{\tt UPV} denotes the valence distribution of up-quarks
$x\Delta u^{\rm val}(x,Q^2)$, {\tt DNV} for down-quarks
$x\Delta d^{\rm val}(x,Q^2)$. {\tt SEA} signifies the sea distribution
$x\Delta q^{\rm sea}(x,Q^2)$. {\tt STR}, {\tt CHM}, {\tt BOT}, and
{\tt TOP} label the distributions for the strange-, charm-, bottom-,
and top-quark $x\Delta q(x,Q^2)$, $q = s, c, b, t$ respectively. Finally
{\tt GLU} marks the gluon distribution $x\Delta g(x,Q^2)$.
\hfill\break
{\bf Remarks:}
{\tt ALTSTI} requires the data file {\tt altsti.dat} which has to be
placed in an appropriate directory. {\tt altsti.dat} is supplied with
this program.

\paragraph{\tt SUBROUTINE POLLO(IFLAG,X,Q2,UVAL,DVAL,GLUE,QBAR,STR)}
\hfill\break
\noindent{\bf Purpose:}
to return $x$ times the polarized parton
distributions evaluated at given $x$ ({\tt X}) and $Q^2$ ({\tt Q2})
according to the parametrization of Gehrmann and Stir\-ling lo~set~a ({\tt
IFLAG=0}), lo~set~b ({\tt IFLAG=1}) or lo~set~c
({\tt IFLAG=2}) \cite{gehrmann}. {\tt UVAL}, {\tt DVAL},
{\tt GLUE}, {\tt QBAR}, {\tt STR} contain $\Delta u_{\rm val}$,
$\Delta d_{\rm val}$, $\Delta g$, $\Delta \bar d=\Delta \bar u$, and
$\delta \bar s$ repectively.
\hfill\break
{\bf Remarks:} {\tt POLLO} requires the subroutines
{\tt RDARRY}, {\tt POLINI} and the data files {\tt gestloa.dat},
{\tt gestlob.dat}, {\tt gestloc.dat}. Like this subroutine they were
written by T.~Gehrmann and W.J.~Stirling and are used here with
permission of the authors. For further informations about this package
see \cite{gehrmann}. The data files have to be
placed in the directory specified in {\tt POLINI}
and are supplied with this program.

\paragraph{\tt SUBROUTINE POLNLO(IFLAG,X,Q2,UVAL,DVAL,GLUE,UBAR,DBAR,STR)}
\hfill\break
\noindent{\bf Purpose:}
to return $x$ times the polarized parton
distributions evaluated at given $x$ ({\tt X}) and $Q^2$ ({\tt Q2})
according to the parametrization of Gehrmann and Stir\-ling nlo~set~a ({\tt
IFLAG=0}), nlo~set~b ({\tt IFLAG=1}) or nlo~set~c
({\tt IFLAG=2}) \cite{gehrmann}. {\tt UVAL}, {\tt DVAL},
{\tt GLUE}, {\tt UBAR}, {\tt DBAR},
{\tt STR} contain $\Delta u_{\rm val}$,
$\Delta d_{\rm val}$, $\Delta g$, $\Delta \bar u$, $\Delta \bar d$, and
$\delta \bar s$ repectively.
\hfill\break
\noindent{\bf Remarks:} {\tt POLNLO} requires the subroutines
{\tt RDARNLO}, {\tt NLOINI} and the data files {\tt gestnloa.dat},
{\tt gestnlob.dat}, {\tt gestnloc.dat}. Like this subroutine they were
written by T.~Gehrmann and W.J.~Stirling and are used here with
permission of the authors. For further informations about this package
see \cite{gehrmann}. The data files have to be
placed in the directory specified in {\tt NLOINI}
and are supplied with this program.

\paragraph{\tt SUBROUTINE PDFSET(PARM,VALUE)}
\hfill\break
\noindent{\bf Purpose:}
new dummy routine for new {\sc Pdflib} interface; has to be removed
if {\sc Pdflib} is used.
\hfill\break
\noindent{\bf Remarks:} This subroutine was taken from {\sc Pythia~5.7}.

\paragraph{\tt SUBROUTINE
STRUCTM(XX,QQ,UPV,DNV,USEA,DSEA,STR,CHM,BOT,TOP,GLU)}
\hfill\break
\noindent{\bf Purpose:}
new dummy routine for new {\sc Pdflib} interface; has to be removed
if {\sc Pdflib} is used.
\noindent{\bf Remarks:} This subroutine was taken from {\sc Pythia~5.7}.

\section{Examples\label{example}}
In the following we give two examples of a main program for a
simulation with {\sc Sphinx} and show the corresponding results. We
considered longitudinal polarized proton-proton scattering in the CMS at
$\sqrt{s}=200\ {\rm GeV}$ and selected the process $qg\rightarrow qg$
for the partonic interaction.
In the first example both beam and target are polarized in
$+z$-direction, in the second example the target spin is reversed.
With regard to the event listings is has to be mentioned that the
displayed format differs from the real because we removed a
few columns such that it fits in this text. In addition the event
listing is cut after the hard interaction (denoted by $\cdots$), i.\
e.\ behind line 8. In
the omitted part there is no polarization information and it has the
same format as the original {\sc Pythia} listing.

The information about the polarization flow is displayed as explained
above right behind the flavour code {\tt KF}. The first row contains the
information about the beam particle. In addition to the {\sc Pythia}
labels the sign `$+$' behind the flavour code for the proton {\tt
KF=2212} denotes the polarization in positive $z$-direction.
Accordingly the sign `$+$' (`$-$') in the second row signifies the
polarization of the target in positive (negative) $z$-direction in the
first (second) example. The third and fourth line represents the
initiators of the initial state shower. In both examples they are
an anti-d-quark and an u-quark with positive helicities for both partons.
During the initial state shower the anti-quark becomes an
gluon with negative helicity, whereas the target side parton
remains a positive helicity u-quark. These partons undergo the hard
interaction the resulting partons of which are displayed in the lines seven
and eight. In the following the final state interaction takes place,
i.\ e.\ the outgoing partons fragment, the unstable produced hadrons decay,
etc.\, as long as only stable particles exist. In the final state
polarization is not traced and consequently there is no polarization
informations about these lines provided. This part of the listing is
then again the same as the corresponding {\sc Pythia} listing.

\subsection{The Main Programs}
\subsubsection{First Example -- parallel polarization}
\begin{verbatim}
C___    Example of a Main Program for event generating
C___    in longitudinal polarized proton-proton-scattering
C___
C___    E.g.: polarized p(+)p(+)-scattering in CMS
C___          at sqrt(s)=200 GeV
C___
C___    This program has to be linked with
C___    the programs SPHINX and JETSET7.3,
C___    the data files ALTSTI.DAT and ROSROA.DAT,
C___    and the CERN-Libraries.


        PROGRAM EXAMPLE

C___    COMMON BLOCKS of SPHINX for event generation

        COMMON/LUDAT1/MSTU(200),PARU(200),MSTJ(200),PARJ(200)
        COMMON/LUJETS/N,K(4000,5),P(4000,5),V(4000,5)
        COMMON/PYSUBS/MSEL,NSEL,MSUB(200),NSUB(200),
     &  KFIN(2,-40:40),CKIN(200)
        COMMON/PYPARS/MSTP(200),PARP(200),MSTI(200),PARI(200)
        COMMON/PYINT5/NGEN(0:200,3),XSEC(0:200,3)

C==============================================================
C       Polarization Set-Up
C==============================================================

C___    Polarized simulation
        MSTP(180)=1

C___    Beam positive polarized
        MSTP(171)=1

C___    Target positive polarized
        MSTP(172)=1

C___    Polarized parton distributions a la Gehrmann Stirling
        MSTP(177)=3

C___    Polarized Initial State Shower
        MSTP(176)=1
C___    with polarized parton distributions
        MSTP(175)=1

C==============================================================

C==============================================================
C       Event Set-Up
C==============================================================

C___    Choice of parton processes ``a la carte''
        MSEL=0

C___    Choice of polarized parton processes ``a la carte''
        NSEL=0

C___    Choice of process: qg --> qg
        MSUB(28)=1

C___    Process 28 polarized
        NSUB(28)=1

C___    Number of generated events
        NEVENT=2000

C___    kinematical cuts
C___    P_T-cut (minimum)
        CKIN(3)=3.
C___    P_T-cut (maximum)
        CKIN(4)=25.

C==============================================================

C==============================================================
C       Start of event generation
C==============================================================

C___    Initialisation

        CALL PYINIT('CMS','p','p',200.)

C___    Loop over events
        DO 100 I=1,NEVENT

C___    Event generation
          CALL PYEVNT

C___    List the first event with spininformation
          IF(I.EQ.1) CALL DPLIST(1)

  100     CONTINUE

C___    Print cross section and histogram
        CALL PYSTAT(1)

        END
\end{verbatim}
\subsubsection{Second Example -- antiparallel polarization}
The second example is constructed by replacing
{\tt MSTP(172)=1} by {\tt MSTP(172)=2} in the first example, i.\ e.\
by switching the spin of the target.
\subsection{The Event Listings}
\subsubsection{First example -- parallel polarization}
\begin{verbatim}
                    SPHINX version 1.1
         **  Last date of change:  18 Nov 1996  **

         The Lund Monte Carlo - PYTHIA version 5.6
         **  Last date of change:   3 Apr 1992  **

         The Lund Monte Carlo - JETSET version 7.3
         **  Last date of change:  10 Mar 1992  **

1******* PYINIT: initialization of PYTHIA routines **********

 ============================================================
 I                                                          I
 I   PYTHIA will be initialized for a p on p collider       I
 I      at    200.000 GeV center-of-mass energy             I
 I                                                          I
 ============================================================

 PYMAXI: summary of differential cross-section maximum search

  ======================================================
  I                                  I                 I
  I ISUB Subprocess name             I  Maximum value  I
  I                                  I                 I
  ======================================================
  I                                  I                 I
  I  28  f + g -> f + g              I    4.7736E+00   I
  I  96  Semihard QCD 2 -> 2         I    1.9598E+02   I
  I                                  I                 I
  ======================================================

 ************** PYINIT: initialization completed *************

                     Event listing (summary)

  I  particle/jet KF    p_x    p_y    p_z       E      m

  1  !p+!       2212+   0.000  0.000  99.996  100.000  0.938
  2  !p+!       2212+   0.000  0.000 -99.996  100.000  0.938
 ============================================================
  3  !d~!         -1+  -0.279 -0.107  12.583   12.587  0.000
  4  !u!           2+  -0.132  0.469  -3.602    3.634  0.000
  5  !g!          21-  -2.214 -1.299   7.340    7.776  0.000
  6  !u!           2+  -0.096  0.342  -2.626    2.649  0.000
  7  !g!          21   -5.168 -1.523   5.182    7.476  0.000
  8  !u!           2    2.857  0.566  -0.467    2.950  0.006
 ============================================================
  ...


1* PYSTAT:  Statistics on Number of Events and Cross-sections *

 ==============================================================
 I                              I                 I           I
 I        Subprocess            I Number of pointsI   Sigma   I
 I                              I                 I           I
 I------------------------------I-----------------I   (mb)    I
 I                              I                 I           I
 I N:o Type                     I Generated Tried I           I
 I                              I                 I           I
 ==============================================================
 I                              I                 I           I
 I  0 All included subprocesses I      2000 13360 I 7.385E-01 I
 I 28 f + g -> f + g            I      2000 13360 I 7.385E-01 I
 I                              I                 I           I
 ==============================================================

 * Fraction of events that fail fragmentation cuts =  0.00000 *
\end{verbatim}
\subsubsection{Second example -- antiparallel polarization}
\begin{verbatim}
                     SPHINX version 1.1
          **  Last date of change:  18 Nov 1996  **


          The Lund Monte Carlo - PYTHIA version 5.6
          **  Last date of change:   3 Apr 1992  **


          The Lund Monte Carlo - JETSET version 7.3
          **  Last date of change:  10 Mar 1992  **

1********* PYINIT: initialization of PYTHIA routines ******


 ==========================================================
 I                                                        I
 I    PYTHIA will be initialized for a p on p collider    I
 I        at    200.000 GeV center-of-mass energy         I
 I                                                        I
 ==========================================================

 PYMAXI: summary of differential cross-section maximum search

 ==========================================================
 I                                      I                 I
 I  ISUB  Subprocess name               I  Maximum value  I
 I                                      I                 I
 ==========================================================
 I                                      I                 I
 I   28   f + g -> f + g                I    4.7851E+00   I
 I   96   Semihard QCD 2 -> 2           I    1.9598E+02   I
 I                                      I                 I
 ==========================================================

 ************ PYINIT: initialization completed **************


                     Event listing (summary)

  I  particle/jet KF    p_x    p_y    p_z       E      m

  1  !p+!       2212+   0.000  0.000  99.996  100.000  0.938
  2  !p+!       2212-   0.000  0.000 -99.996  100.000  0.938
 ============================================================
  3  !d~!         -1+  -0.279 -0.107  12.583   12.587  0.000
  4  !u!           2+  -0.132  0.469  -3.602    3.634  0.000
  5  !g!          21-  -2.214 -1.299   7.340    7.776  0.000
  6  !u!           2+  -0.096  0.342  -2.626    2.649  0.000
  7  !g!          21   -5.168 -1.523   5.182    7.476  0.000
  8  !u!           2    2.857  0.566  -0.467    2.950  0.006
 ============================================================
  ...



1* PYSTAT:  Statistics on Number of Events and Cross-sections *

 ==============================================================
 I                              I                 I           I
 I         Subprocess           I Number of pointsI   Sigma   I
 I                              I                 I           I
 I------------------------------I-----------------I   (mb)    I
 I                              I                 I           I
 I N:o Type                     I Generated Tried I           I
 I                              I                 I           I
 ==============================================================
 I                              I                 I           I
 I  0 All included subprocesses I 2000      12655 I 7.361E-01 I
 I 28 f + g -> f + g            I 2000      12655 I 7.361E-01 I
 I                              I                 I           I
 ==============================================================

 * Fraction of events that fail fragmentation cuts =  0.00000 *

\end{verbatim}
\section*{Acknowledgements}
We all thank T.\ Sj\"ostrand very much for his help and encouragement
in understanding the details of the {\sc Pythia} code.
A.\ S.\ thanks the MPI f\"ur Kernphysik for its support.
The work of
L.\ M.\  was supported in part by KBN under grant 2~P302~143~06. This work
was supported also by DFG (Scha 458/3-1) and BMFT-KBN (Projekt X081.92)

\end{document}